\documentclass[10pt,journal,twocolumn]{IEEEtran}
\usepackage{amssymb,amsmath}

\newtheorem{Definition}{Definition}

\usepackage{cite}
\usepackage{graphics}
\usepackage{graphicx}
\usepackage{subfigure}
\begin{document}

\title{Distributed Space-Time Coding for Full-Duplex Asynchronous Cooperative
Communications\thanks{Liu and Zhang are with the State Key Laboratory of Integrated Service Network, Xidian University, Xi'an, 710071, China (\{yliu, hlzhang\}@xidian.edu.cn). Their research was supported in part by the Important National Science \& Technology Specific Projects (2011ZX03003-001-04, 2012ZX03003012-003), the 111 Project (B08038), and the Natural Science Foundation (61072069), China. Xia is with the Department of Electrical and Computer Engineering, University of Delaware, Newark, DE 19716, USA (xxia@ee.udel.edu), and also the Chonbuk National University, Jeonju, South Korea. His research was supported in part by the National Science Foundation (NSF) under Grant CCF-0964500, the Air Force Office of Scientific Research (AFOSR) under Grant FA9550-12-1-0055, and the World Class University
(WCU) Program, National Research Foundation, South Korea.}
}

\author{Yi Liu, Xiang-Gen Xia, and Hailin Zhang}

\maketitle

\begin{abstract}
In this paper, we propose
two distributed linear convolutional space-time coding (DLC-STC) schemes
 for  full-duplex (FD) asynchronous cooperative communications.
The DLC-STC Scheme 1 is for the case of the complete loop channel cancellation, which
achieves the full asynchronous cooperative diversity.
The DLC-STC Scheme 2 is for the case of the partial loop channel cancellation and amplifying, where some loop signals are used
as the self-coding instead of treated as interference to be directly cancelled. We show this scheme can achieve full asynchronous
cooperative diversity. We then evaluate the performance of the two schemes
when loop channel information is not accurate and present an amplifying factor control method for the DLC-STC Scheme 2
to improve its performance with inaccurate loop channel information. Simulation results show that the DLC-STC Scheme 1 outperforms the DLC-STC
Scheme 2 and the delay diversity scheme if perfect or high quality loop channel information is available at the relay,
while the DLC-STC Scheme 2 achieves better performance if the loop channel information is imperfect.

\end{abstract}

\begin{IEEEkeywords}
Asynchronous cooperative diversity, cooperative communications, distributed space-time coding, full-duplex transmission.
\end{IEEEkeywords}

\IEEEpeerreviewmaketitle

\section{Introduction}

Cooperative communications has attracted significant attention
in the last decade, see for example,
\cite{Sendonaris01}--\cite{5683902},
where relay nodes may process and transmit the information they receive.
The processing in relay nodes is usually classified into amplify-and-forward (AF) and decode-and-forward (DF) protocols. Another classification for
relay nodes is  full-duplex (FD) or half-duplex (HD) mode.
 In  the HD mode, a relay is restricted to receive and transmit on orthogonal (in time or in frequency) channels, which may not be spectrally efficient.
 In the FD mode~\cite{5089955,5161790,5470111,5961159,4917875}, on the other hand, a relay only requires one channel for the end-to-end transmission. As a result, in general,
FD cooperative protocols achieve a higher capacity than HD cooperative protocols~\cite{1499041}. However, the FD mode introduces loop (self)
 interference due to the signal leakage between the relay output and input, which may be a serious problem for small portable devices and therefore is
the reason why most research in cooperative communications is on the HD mode.

Attracted by the high spectrum efficiency,
the feasibility of FD relaying in the presence of loop interference has been
recently studied. In~\cite{5089955}, an AF protocol is considered for
FD mode cooperative communications and a smart relay gain control strategy is proposed with the effect of residual loop interference.
In~\cite{5161790}, the break-even loop interference level is evaluated and it shows that the FD mode offers capacity improvement over the HD mode.
In~\cite{5470111},  MIMO relay case with FD mode is investigated and
it considers the design of linear receivers and transmit filters for the relays
to improve the quality of the useful signal while minimizing the effect of the loop interference.
Hybrid techniques that switch between full-duplex and half-duplex relay modes are developed in~\cite{5961159} for maximizing instantaneous and average spectrum efficiency. An FD mode cooperative communication system with one relay and
one direct link is considered in~\cite{4917875}. It provides the closed-form expressions for outage probability
by taking into account of practical constraints such as interference
due to frequency reuse and signal leakage between the relay
transmitter and receiver.
All these literatures regard the signals from loop channels at the relays as interferences and try to remove them as much as possible.

In this paper, we propose two distributed linear convolutional space-time coding (DLC-STC) schemes
 for  full-duplex asynchronous cooperative communications.
We first propose a DLC-STC scheme (called DLC-STC Scheme 1) for the case when
the  loop channel interference is cancelled completely, where
the DLC-STC obtained in~\cite{Guo} is directly used.  This scheme
achieves the full asynchronous cooperative diversity.
 We then propose a DLC-STC scheme (called DLC-STC Scheme 2)  for the case when the loop channel
interference is intended not to be cancelled completely, where some loop signals are used as the self-coding. We show that this scheme can achieve full asynchronous cooperative diversity. We also evaluate the effect of the loop channel information error and show that with the amplifying factor control method, the DLC-STC Scheme 2 has better tolerance to the loop
channel information error than the DLC-STC Scheme 1 and the delay diversity scheme. The idea of amplifying factor control method is the same as that in~\cite{5089955}, which is to adjust the relay transmission power according to the loop channel information error. However, there are two  main differences. One is the amplifying factor control in~\cite{5089955} regards the signal from the direct link as noise while in this paper, the signal from the direct link is treated as useful signal. The other is for the proposed two DLC-STC schemes, the useful signal sent by the relay is made up of several consecutive symbols while for~\cite{5089955}, the useful signal sent by the relay is only one symbol.

This paper is organized as follows. In Section \ref{sec2},
we  formulate the system and signal models for two-hop FD relay links.
In Section \ref{sec3}, we present two construction methods of DLC-STC for full
duplex asynchronous cooperative communications. We also
 investigate the diversity
of the proposed schemes when there are different delays at the links. In Section \ref{sec5}, we  analyze the impact of loop channel information errors and present an amplifying factor control method for the DLC-STC Scheme 2.
In Section \ref{sec6}, we present some simulation results to evaluate
the performance of the proposed schemes. Finally,
in Section \ref{sec7}, we conclude this paper.

\section{System model}\label{sec2}

\begin{figure}[htbp]
\centering
\includegraphics{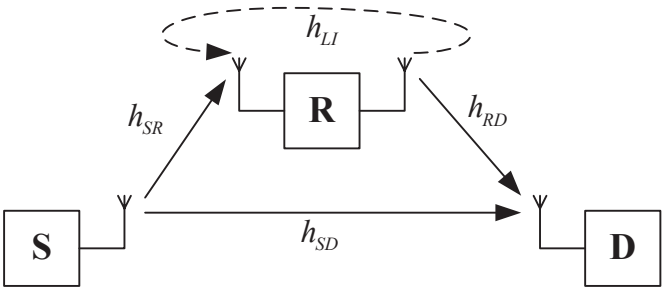}
\caption{\label{fig1}Full-duplex two-hop relay network with direct link used in \cite{4917875}.}
\end{figure}
The system model  in this paper follows that used in ~\cite{4917875}. And the signal model of FD mode relay networks follows what is used  in~\cite{5089955,5470111,4917875}. First, let us briefly recall the system and signal models
studied in ~\cite{5089955,5470111,4917875}.

Consider a cooperative relay network shown in Fig.\ref{fig1}, where there is one relay node between the source node and the destination node~\cite{4917875}. The source node communicates with the destination node via both the relay node and the direct link. The relay node receives and transmits signals with the same frequency at the same time. So there is a loop link between the transmitter and the receiver of the relay node. The AF protocol is adopted at the relay. There are four physical channels, namely source to relay channel $h_{SR}$, relay to destination channel $h_{RD}$, relay loop channel $h_{LI}$, and source to destination channel $h_{SD}$. All of them are assumed to be quasi-static flat fading and i.i.d.,
 and have the distribution of $\mathcal{CN}(0,1)$. The received signal $r(i)$ and the transmitted signal $t(i)$ at the relay at time $i$ are
\begin{eqnarray}
r(i)&=&h_{SR}x(i)+h_{LI}t(i)+n_R(i) \label{equ1}\\
\nonumber t(i)&=&\beta r(i-1)\\&=&\beta \sum\limits_{j=1}^{\infty}(h_{LI}\beta)^{j-1}[h_{SR}x(i-j)+{n}_R(i-j)]\label{equ2},
\end{eqnarray}
where $\beta$ is the amplifying factor to normalize the transmission power at the relay, $x(i)$ is the transmitted signal by the source node
with normalized power $E_s=E[|x(i)|^2]=1$, and $n_R(i)$ is the additive $\mathcal{CN}(0,\sigma_R^2)$ noise at the receiver of the relay.
Suppose the maximum relative delay between the direct link and the relay link is $\tau_{max}$. The signal received at the destination from the relay is
\begin{equation}\label{equ3}
\begin{array}{rcl}
y(i)&=&h_{RD}t(i)+h_{SD}x(i+\tau)+n_D(i)\\
&=&\beta h_{RD}\sum\limits_{j=1}^{\infty}(h_{LI}\beta)^{j-1}[h_{SR}x(i-j)+{n}_R(i-j)]\\
&+&h_{SD}x(i+\tau)+n_D(i),
\end{array}
\end{equation}
where $|\tau|\leq \tau_{max}$ and $n_D(i)\sim\mathcal{CN}(0,\sigma_D^2)$ is the additive Gaussian noise at the destination node.

Generally, the signal from the loop link is regarded as interference and tried to be removed from the received signal at the relay node,
which is called loop cancellation. The relay subtracts an estimate of the loop interference from its input and then amplifies the result
by factor $\beta>0$~\cite{5089955}. Thus, we can reformulate (\ref{equ2}) as
\begin{equation}
t(i)=\beta [r(i-1)-wt(i-1)]  \label{equ4}
\end{equation}
where $w$ is the cancellation filter coefficient. The transmission power at the relay is normalized to be 1, that is, $\mathcal{E}\{|t(i)|^2\}=1$.

By recursive substitution of (\ref{equ1}) and (\ref{equ4}) we obtain
\begin{equation}\label{equ5}
t(i)=\beta \sum\limits_{j=1}^{\infty}(\hat{h}_{LI}\beta)^{j-1}[h_{SR}x(i-j)+{n}_R(i-j)]
\end{equation}
where $\hat{h}_{LI}=h_{LI}-w$ is the residual loop channel. The loop interference cancellation coefficient $w$ can be determined by
any existing adaptive filtering, or pilot-based channel estimation method~\cite{5089955}. One special case called complete loop interference cancellation is that
the relay  ideally use $w=h_{LI}$, which results in no residual loop interference.

Finally, the signal received by the destination from the relay is
\begin{equation}\label{equ6}
\begin{array}{rcl}
y(i)&=&h_{RD}\beta h_{SR}x(i-1)+h_{SD}x(i+\tau)\\
&+&\sum\limits_{j=2}^{\infty}(\hat{h}_{LI}\beta)^{j-1}[h_{SR}x(i-j)+{n}_R(i-j)]\\&+&h_{RD}\beta {n}_R(i-1)+n_D(i).
\end{array}
\end{equation}

Usually the first two terms in the right hand side of (\ref{equ6})
are the desired signals and the third term  is regarded as loop interference
that should be cancelled as much as possible by minimizing $\hat{h}_{LI}^{(k)},k=0,1,\ldots,r-1$,~\cite{5089955,5470111}. In this paper, we will propose
two methods to design DLC-STC for the  relay link and the direct link. One is
for the case when  the loop channel is completely
cancelled.
The other is to make use of the loop link interference
partially by controlling the parameter $\beta$ in  a proper way
to better exploit the spatial diversity.

\section{Construction of DLC-STC for Full-Duplex Asynchronous Cooperative Communications}\label{sec3}
 Our goal is to design a DLC-STC achieving full diversity with the loop interference channel at the relay. In this section, we will show two schemes
 for constructing the DLC-STC. One is to cancel the loop channel completely and then code with a predesigned coding matrix. The other is to cancel the  loop channel partially and code by making use of the residual loop channel at the relay.
Note that although we have only one relay, we can treat the direct link as a special 'relay' link.

\subsection{Scheme 1: Complete Loop Channel Cancellation and Coding}
\begin{figure}[htbp]
\centering
\includegraphics{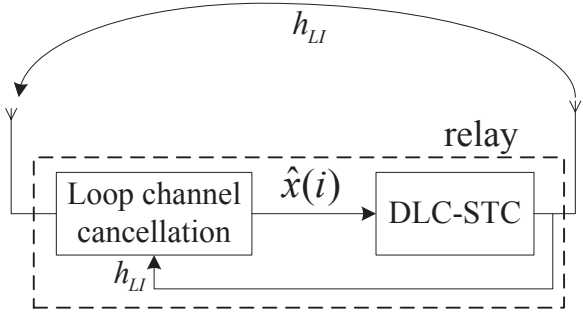}
\caption{\label{scheme1}Complete loop channel cancellation DLC-STC scheme at the relay for full-duplex cooperative communications.}
\end{figure}

As shown in Fig.\ref{scheme1}, if the loop channel is known at the relay, the signal from the loop channel can be removed and the signal from the source node can be estimated as

\begin{equation}\label{equ701}
\hat{x}(i)=r(i)-h_{LI}t(i)=h_{SR}x(i)+n_R(i),
\end{equation}
which then becomes the same as the HD mode.
Thus, we may apply the existing schemes for HD model to the above FD mode.
One of such schemes we are interested is the DLC-STC scheme
proposed in \cite{Guo} as an AF protocol for an HD relay network
and it achieves the full asynchronous cooperative diversity
where the synchronization between the links is not required.
We next briefly review the DLC-STC scheme in \cite{Guo} and this concept
will be also used in  the second scheme we want to propose in the next subsection.

Assume that the information sequence is $\mathbf {s}=[s_1, s_2,\ldots,s_{l}]^T$. At the $k$th relay (There are two relays in our case since the direct link is regarded as a special relay.),
it is
transformed into $\mathbf{c}_k$ by a
preassigned transform matrix $\mathbf{A}_k \in \mathbb{C}^{q\times l}$, i.e., $\mathbf{c}_k=\mathbf{A}_k\mathbf{s}$. All the two relays form
the space-time code matrix $\mathbf{C}=[\mathbf{c}_0,\mathbf{c}_1]^T$, which is called distributed linear dispersion
space-time code. If the transform matrix $\mathbf{A}_k=(a_{i,j})$
satisfies the following condition:
\begin{equation}\label{9}
  a_{i,j}=\left\{\begin{array}{ll} v_{i-j+1} & 0\leq i-j<b; \\
  0, & \textrm{otherwise,}
  \end{array} \right.
\end{equation}
for a sequence $\mathbf{v}_k=[v_1, v_2, \ldots, v_{b}]^T$ of length $b$,
$\mathbf{A}_k\in \mathbb{C}^{q\times l}$ where $q=b+l-1$, is called a convolution matrix.
If all the matrices $\mathbf{A}_k, 0\leq k \leq 1$, are convolution matrices, the space-time code $\mathbf{C}$ is called a distributed linear convolutional space-time
code (DLC-STC). Assume that the
transmission delay for the $k$th relay link $\tau_k, k=0,1$, is an integer multiple of $t_s$, where $t_s$ is the information symbol
period.
To deal with the timing errors among the relays is to protect
the coded symbol sequences/frames with guard intervals by zero padding. The zero padding length $\tau_{max}$ is assumed not smaller than the maximum
of all the delays $\tau_k$.
Then, the received signal $\mathbf y$ at the destination
can be written as
\begin{equation}\label{10}
\mathbf y=\mathbf {hC}_\triangle +\mathbf n
\end{equation}
where $\mathbf h$ is the $1\times 2$ channel matrix and $\mathbf n$ is the noise, $\mathbf {C}_\triangle \in \mathbb C^{r\times
(l+ \tau_{max})}$ is a time shifted version of the zero padded code matrix, which has the follow form:
\begin{equation}
\left[ \begin{array}{c} \left[\mathbf 0_{\tau_0}\,\,\,\,\,\,\mathbf{c}_0^T\,\,\,\,\,\,\mathbf 0_{\tau_{max}-\tau_0}\right]\\
\left[\mathbf 0_{\tau_1}\,\,\,\,\,\,\mathbf c_1^T\,\,\,\,\,\,\mathbf 0_{\tau_{max}-\tau_1}\right]
\end{array}\right]
\end{equation}
where $\mathbf 0_n$ denotes a $1\times n$ all-zero vector. The channel state information (CSI) such as channel coefficients and delay profiles
is assumed known at the destination node. Thus, the receive signal model
(\ref{10}) can be effectively viewed as a result of the codeword $\mathbf {C}_\triangle$
sent through the channel $\mathbf{h}$ and $\mathbf {C}_\triangle$ is called an effective code. To achieve the full cooperative diversity, the effective code $\mathbf {C}_\triangle$ needs to have the full rank
property.
The problem is that the delay profile $\triangle$ may vary from time to time due to the dynamics of a network
topology and it is not known at the relay nodes. The question is how to design the DLC-STC such that the effective code $\mathbf {C}_\triangle$
has the full rank property for all delay profile $\triangle$.
This question has been studied in \cite{4155128, Shang001, Guo}
by using shift-full-rank (SFR) matrices as generator matrices, which is recalled below.

\begin{Definition}
\label{def1}
Let $\mathbf{M}=[\mathbf{m}_0^T,\mathbf{m}_1^T,\cdots,\mathbf{m}_{r-1}^T]^T\in \mathbb{C}^{
r\times b}$, where $\mathbf{m}_k$ are row vectors of $\mathbf{M}$.
If for  any delay profile $\Delta=\{\tau_0, \tau_1,\cdots,\tau_{r-1}\}$,
 the following matrix
\begin{equation}
\left[ \begin{array}{c} \left[ \mathbf 0_{\tau_0}\,\,\,\,\,\,\,\,\,\,\,\,\mathbf{m}_0\,\,\,\,\,\,\,\,\,\,\,\,\mathbf 0_{\tau_{max}-\tau_0}\right]\\
\left[\mathbf 0_{\tau_1}\,\,\,\,\,\,\,\,\,\,\,\,\mathbf{m}_1\,\,\,\,\,\,\,\,\,\,\,\,\mathbf 0_{\tau_{max}-\tau_1}\right]\\
\vdots\\
\left[\mathbf 0_{\tau_{r-1}}\,\,\,\,\,\,\mathbf{m}_{r-1}\,\,\,\,\,\, \mathbf 0_{\tau_{max}-\tau_{r-1}}\right]
\end{array}\right]
\end{equation}
is always a full row rank matrix, then we call $\mathbf{M}$ a {\it shift-full rank}
(SFR) matrix,
where $\tau_{max}$ is the maximum delay among all $\tau_k$.
\end{Definition}

It is shown in \cite{4155128, Shang001, Guo} that, if the DLC-STC is generated from an SFR matrix, it can achieve asynchronous full diversity.

For the case in this paper, the direct link can be regarded as a special 'relay' forming the first row of the generator matrix $\mathbf{M}$ and the relay link forms the other row of $\mathbf{M}$ and thus
\begin{equation}\label{M2}
\mathbf{M}=\left[\begin{array}{cccc}1&0&\ldots&0\\ m_1&m_2&\ldots&m_b\end{array}\right].
\end{equation}
The transmission power at the relay should be normalized to 1, so we have $\sum_{j=1}^b|m_j|^2=1$. It is obvious that the condition for the generator matrix $\mathbf{M}$ to be an SFR matrix is $|m_j|^2\neq1, 1\leq j\leq b$. With such an SFR matrix $\mathbf{M}$, the transform matrix
$\mathbf{A}$ for the relay is constructed by
forming a convolution matrix~(\ref{9}) using the second row vector
of the matrix $m_j$ as $a_{j}$, i.e., $v_j=m_{j}$, and
the DLC-STC coding at the relay is:
\begin{equation}\label{equ702}
t(i)=\sum\limits_{j=1}^b m_{j} \hat{x}(i-j),
\end{equation}
where $ \hat{x}(i)$ is the signal after the self-loop interference
cancellation in~(\ref{equ701}). It is shown in \cite{Guo}
that this DLC-STC achieves full cooperative diversity no matter how a delay
profile $\Delta$ from the relay is, i.e., it achieves the full asynchronous cooperative diversity, with the maximum-likelihood (ML) or the MMSE-DFE receiver.

It should be noted that the selection of $m_j, 1\leq j\leq b$, is not unique to ensure $\mathbf{M}$ an SFR matrix. In the simulation section, we use $m_j=1/\sqrt{b}, 1\leq j\leq b$, for Scheme 1.

Specially, if $m_i= 1$ and $m_j= 0, 1\leq j\leq b, j\neq i$, the code becomes the delay diversity code~\cite{tarokh98,997188}, which cannot achieve full diversity in asynchronous systems.

\subsection{Scheme 2: Partial Loop Channel Cancellation and Amplifying}
 \begin{figure}[htbp]
\centering
\includegraphics{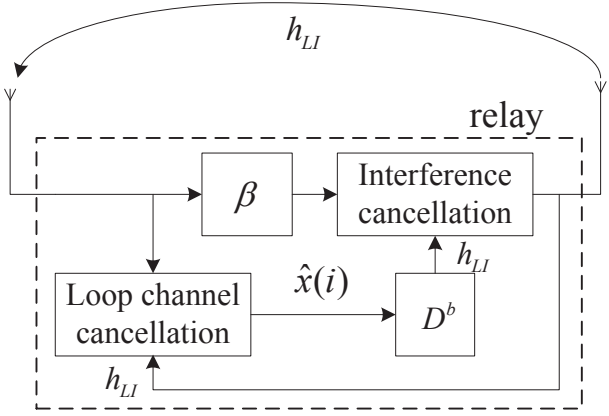}
\caption{\label{scheme2}Partial loop channel cancellation and amplifying DLC-STC scheme at the relay for full-duplex cooperative communications.}
\end{figure}
In this scheme, the signal from the loop channel is removed partially and then amplified by a factor before the transmission. The process is
shown in Fig.\ref{scheme2}. The signal from the source node is also estimated first. However, the difference is that the estimated signal is not used for coding but used for interference cancellation after delayed by $b$ sample periods (denoted as $D^b$ in the figure). Amplifying and interference cancellation are done to the signal before its retransmission. We will show that  the signals from the relay link and the direct link form a DLC-STC .

The same as Scheme 1, the relay in Scheme 2 needs to know the
relay loop channel $h_{LI}$. On the other hand, different from Scheme 1, not all the signal from the loop channel but the signal after a certain times of loops is cancelled and
the remaining loop signal is used as a self-code as shown below.


Suppose $b$ consecutive  symbols are  to be coded. In~(\ref{equ2}), by letting $i=b$ and considering $\{x(i)=0,n_R(i)=0,i\leq -1\}$, we can obtain the transmit signal at the relay at time slot $b$ as
\begin{equation}\label{equ11-00}
t(b)=\beta \sum\limits_{j=1}^{b}(h_{LI}\beta)^{j-1}[h_{SR}x(b-j)+{n}_R(b-j)].
\end{equation}
Substituting~(\ref{equ11-00}) into~(\ref{equ1}), the received signal at the relay at time slot $b$ can be written as
\begin{equation}\label{equ11}
\begin{array}{rcl}
r(b)&=& h_{SR}x(b)+h_{LI}t(b)+{n}_R(b)\\
      &=&\sum\limits_{j=1}^{b}(h_{LI}\beta)^{j-1}h_{SR}x(b-j+1)\\
      &+&(h_{LI}\beta)^{b}[h_{SR}x(0)+n_R(0)]\\
      &+&\sum\limits_{j=1}^{b}(h_{LI}\beta)^{j-1}n_R(b-j+1).
\end{array}
\end{equation}
The first term in the right hand side of (\ref{equ11}) is the desired transmit signal including
consecutive  $b$ symbols while the second and third terms are
interference and noise from the loop channel. Notice that the second term in (\ref{equ11}) can be written as
$$(h_{LI}\beta)^{b}[h_{SR}x(0)+n_R(0)]=(h_{LI}\beta)^{b}\hat{x}(0).$$
 At time slot $b+1$, since the relay has obtained
 the estimated signal $\hat{x}(0)$ as in (\ref{equ701}), the interference of the second term in (\ref{equ11})
can be cancelled. Then the transmit signal can be written as
\begin{equation}\label{equ11-20}
\begin{array}{rcl}
t(b+1)&=&\beta [r(b)-(h_{LI}\beta)^{b}\hat{x}(0)]\\
&=&\beta \sum\limits_{j=1}^{b}(h_{LI}\beta)^{j-1}h_{SR}x(b+1-j)\\
&+&\beta \sum\limits_{j=1}^{b}(h_{LI}\beta)^{j-1}{n}_R(b+1-j).
\end{array}
\end{equation}
If the cancellation process is done continuously
in terms of $k$ in the  time index $b+k$,
we can obtain a general expression as
\begin{equation}\label{equ11-21}
\begin{array}{rcl}
t(b+k)&=&\beta [r(b+k-1)-(h_{LI}\beta)^{b}\hat{x}(k-1)]\\
&=&\beta \sum\limits_{j=1}^{b}(h_{LI}\beta)^{j-1}h_{SR}x(b+k-j)\\
&+&\beta \sum\limits_{j=1}^{b}(h_{LI}\beta)^{j-1}{n}_R(b+k-j).
\end{array}
\end{equation}
Letting $i=b+k$, (\ref{equ11-21}) can be simplified as
\begin{equation}\label{equ11-2}
\begin{array}{rcl}
t(i)&=&\beta [r(i-1)-(h_{LI}\beta)^{b}\hat{x}(i-b-1)]\\
&=&\beta \sum\limits_{j=1}^{b}(h_{LI}\beta)^{j-1}h_{SR}x(i-j)\\
&+&\beta \sum\limits_{j=1}^{b}(h_{LI}\beta)^{j-1}{n}_R(i-j).
\end{array}
\end{equation}
Let $q(i)=\beta(h_{LI}\beta)^{i-1}h_{SR},1\leq i \leq b$, and then the first
term in the right hand side of ~(\ref{equ11-2}) can be written as the convolution
between $q(i)$ and $x(i)$ as follows:
\begin{equation}\label{equ13}
\beta\sum\limits_{j=1}^{b}(h_{LI}\beta)^{j-1}h_{SR}x(i-j)=q(i)\ast x(i),
\end{equation}
where $0<\beta<\frac{1}{|h_{LI}|}$ is the amplify parameter controlling the relay transmission power as
\begin{equation}\label{equ13-2}
\begin{array}{rcl}
\mathcal{E}\{\sum_{i=1}^b|q(i)|^2\}&=&\mathcal{E}\{|h_{SR}|^2\}\sum_{i=1}^b|\beta(h_{LI}\beta)|^{i-1}\\
&=&\sum_{i=1}^b|\beta(h_{LI}\beta)|^{i-1}\leq1.
\end{array}
\end{equation}

In (\ref{equ13-2}), $b$ determines the constraint length of the convolution code.
To ensure the effective coding matrix of full row rank
to achieve the full cooperative diversity,
$b$ should be no less than the number of independent links, which is  $2$ in the current case. The function
$q(i)=\beta(h_{LI}\beta)^{i-1}h_{SR},1\leq i \leq b$,
 is determined after $\beta$ is selected to satisfy  (\ref{equ13-2}).
If we combine the relay link
 with the above encoding (\ref{equ13})
and the direct link,
the signal at the destination can be thought of as
a DLC-STC with the following generator matrix:
\begin{equation}\label{Mscheme2}
 \mathbf{M}=\left[\begin{array}{cccc}1&0&\cdots&0\\ \beta&\beta(h_{LI}\beta)&\cdots&\beta(h_{LI}\beta)^{b-1}\end{array}\right].
\end{equation}
Note that $h_{LI}$ is the self loop channel coefficient and it does not have a small value. Thus, when  $b\geq2$, it is easy to see that
the above generator matrix $\mathbf{M}$ in (\ref{Mscheme2}) is an SFR matrix.
Furthermore, in order to normalize the mean transmission power at the relay,
the norm of the second row vector in $\mathbf{M}$ in (\ref{Mscheme2})
is normalized to $1$. Thus, although the coefficient $h_{LI}$ in the code generator
matrix $\mathbf{M}$ in (\ref{Mscheme2}) is random, the code property should be similar to
those of deterministic coefficents
and therefore the
DLC-STC in this scheme achieves full asynchronous cooperative diversity.
This will be illustrated later  by simulations.

The difference of the DLC-STC in Scheme 2 with that in Scheme 1 is
that the one in Scheme 2 is automatically
inherited from the loop channel by only maintaining a few signals from the
loop channel, while the one in Scheme 1 is re-encoding after the loop channel interference is completely cancelled. Their performance comparison will be given in the simulation section.

\section{Impact of Loop Channel Information Errors}\label{sec5}
In this section, we will investigate the performance of the proposed two schemes when the  loop channel information has errors.
Suppose the loop channel is $h_{LI}$
and the relay estimates the channel information with an error $\Delta h\in \mathcal{CN}(0,\sigma_h^2)$.
It is clear that if the
loop channel error is large,
the system performance will not be acceptable. Thus,
it is reasonable to assume the loop channel error
$|\Delta h|\ll 1$. The loop channel information obtained by the relay is
\begin{equation}\label{equ21}
\bar{h}_{LI}=h_{LI}-\Delta h.
\end{equation}
Then, the estimated signal at the relay in (\ref{equ701})  becomes
\begin{equation}\label{equ22}
\hat{x}(i)=r(i)-\bar{h}_{LI}t(i)=h_{SR}x(i)+\Delta ht(i)+n_R(i).
\end{equation}

\subsection{SINR of Scheme 1}\label{sec5-b}
For the proposed DLC-STC Scheme 1, substituting (\ref{equ22}) to (\ref{equ701}), we obtain
\begin{equation}\label{equ28}
\begin{array}{rcl}
t(i)&=&\dfrac{1}{h_{SR}}\sum\limits_{j=1}^b m_j [h_{SR}{x}(i-j)+n_R(i-j)]\\&+&\dfrac{\Delta h}{h_{SR}}\sum\limits_{j=1}^b m_j t(i-j),
\end{array}
\end{equation}
where the DLC-STC generator coefficients $m_i$ are assumed all reals.
The last term in the right hand side of  (\ref{equ28}) is the interference produced by the loop channel error. Due to the assumption that
$|\Delta h|\ll 1$,
 we may only consider
the interference produced by $\Delta h$ and neglect that from the terms of
$(\Delta h)^p(p\geq2)$. Then, the interference due to the loop channel error
in DLC-STC Scheme 1 is
\begin{equation}\label{equ28-2}
\begin{array}{rcl}
t_{\Delta h}(i)=\dfrac{\Delta h}{h_{SR}^2}\sum\limits_{j=2}^{2b}\sum\limits_{\substack{u+v=j\\1\leq u,v\leq b }}m_um_v[h_{SR}{x}(i-j)+n_R(i-j)].
\end{array}
\end{equation}
Considering $x(i), i=0,1,2,\ldots$ and $n_R(i), i=0,1,2,\ldots$ are independent, we get the power of interference in DLC-STC Scheme 1 as follows:
\begin{equation}\label{equ28-3}
P_{IS1}=\frac{\sigma_h^2}{|h_{SR}|^4}\sum\limits_{j=2}^{2b}|\sum\limits_{\substack{u+v=j\\1\leq u,v\leq b }}m_um_v|^2(|h_{SR}|^2+\sigma_R^2).
\end{equation}
From (\ref{equ3}) and (\ref{equ28}), we obtain the SINR at the destination node as
\begin{equation}\label{equ28-4}
\gamma_{S1}=\frac{|h_{RD}|^2\sum\limits_{j=1}^{b} |m_j|^2+|h_{SD}|^2}{|h_{RD}|^2\left(\frac{\sigma_R^2}{|h_{SR}|^2}\sum\limits_{j=1}^{b} |m_j|^2+P_{IS1}\right)+\sigma_D^2}.
\end{equation}
Since $\sum\limits_{j=1}^{b} |m_j|^2=1$, we get
\begin{equation}\label{equ28-5}
\gamma_{S1}=\frac{1+\frac{|h_{SD}|^2}{|h_{RD}|^2}}{\frac{\sigma_R^2}{|h_{SR}|^2}+P_{IS1}+\frac{\sigma_D^2}{|h_{RD}|^2}}.
\end{equation}

When  $m_i= 1$ and $m_j= 0, 1\leq j\leq b, j\neq i$, the above DLC-STC
is reduced to the delay diversity code \cite{tarokh98,997188}. In this case, the SINR becomes:
\begin{equation}\label{equ26}
P_{IDD}=\frac{\sigma_h^2}{|h_{SR}|^2}(1+\frac{\sigma_R^2}{|h_{SR}|^2})
\end{equation}
\begin{equation}\label{equ26-0}
\gamma_{DD}=\frac{1+\frac{|h_{SD}|^2}{|h_{RD}|^2}}{\frac{\sigma_R^2}{|h_{SR}|^2}+P_{IDD}+\frac{\sigma_D^2}{|h_{RD}|^2}}.
\end{equation}

Comparing with the delay diversity code, the proposed DLC-STC with
an SFR generating matrix has lower SINR when the loop channel
information is not perfect as shown below.

From (\ref{equ28-3}), we have
\begin{equation}\label{equ29}
\begin{array}{rcl}
P_{IS1}&=&\dfrac{\sigma_h^2}{|h_{SR}|^4}\sum\limits_{j=2}^{2b}|\sum\limits_{\substack{u+v=j\\1\leq u,v\leq b }}m_um_v|^2(|h_{SR}|^2+\sigma_R^2)\\
&\stackrel{\text{(i)}}{\geq}&\dfrac{\sigma_h^2}{|h_{SR}|^4}\sum\limits_{j=2}^{2b}\sum\limits_{\substack{u+v=j\\1\leq u,v\leq b }}|m_um_v|^2(|h_{SR}|^2+\sigma_R^2)\\
&=&\dfrac{\sigma_h^2}{|h_{SR}|^4}(\sum\limits_{i=1}^{b}|m_i|^2)^2(|h_{SR}|^2+\sigma_R^2),
\end{array}
\end{equation}
where the inequality (i) can be proved by expanding the summations on both sides of the inequality to obtain the sum of squares when $b\leq 4$ and is conjectured
true for all $b$ since it is testified by numerical results.
Since $\sum\limits_{j=1}^{b} |m_j|^2=1$, we get the power of interference
\begin{equation}\label{equ29-2}
P_{IS1}\geq\frac{\sigma_h^2}{|h_{SR}|^2}(1+\frac{\sigma_R^2}{|h_{SR}|^2})=P_{IDD}.
\end{equation}
Thus,
$\gamma_{S1}\leq \gamma_{DD}$.

The above result implies that the loop channel error has more
effect on the DLC-STC  Scheme 1 with an SFR generating matrix
than that of the delay diversity code.

\subsection{SINR of Scheme 2}\label{sec5-c}
For the proposed DLC-STC Scheme 2, if there exists an error $\Delta h$ in the loop channel information $\bar{h}_{LI}=h_{LI}-\Delta h$,
the transmitted signal at the relay in (\ref{equ11-2}) becomes~(\ref{equ30}) at the top of the next page,
where $t_{\Delta h}(i)$ represents all the terms produced by $\Delta h$.

\begin{figure*}[!t]
\normalsize
\begin{equation}\label{equ30}
\begin{array}{rcl}
t(i)&=&\beta [r(i)-(\bar{h}_{LI}\beta)^{b}\hat{x}(i-b-1)]\\
&=&\beta \sum\limits_{j=1}^{b+1}((\bar{h}_{LI}+\Delta h)\beta)^{j-1}[h_{SR}x(i-j)+{n}_R(i-j)]\\
&-&\beta^{b+1}\bar{h}_{LI}^b[h_{SR}x(i-b-1)+{n}_R(i-b-1)+\Delta h t(i-b-1)]\\
&=&\beta \sum\limits_{j=1}^{b}(\bar{h}_{LI}\beta)^{j-1}[h_{SR}x(i-j)+{n}_R(i-j)]+t_{\Delta h}(i)
\end{array}
\end{equation}
\begin{equation}\label{equ31}
\begin{array}{rcl}
t_{\Delta h}(i)&=&\beta \sum\limits_{j=2}^{b+1}C_{j-1}^1\bar{h}_{LI}^{j-2}\Delta h\beta^{j-1}[h_{SR}x(i-j)+{n}_R(i-j)]\\
&-&\beta^{b+2}\bar{h}_{LI}^b\Delta h \sum\limits_{j=1}^{b}(\bar{h}_{LI}\beta)^{j-1}[h_{SR}x(i-b-1-j)+{n}_R(i-b-1-j)]
\end{array}
\end{equation}
\begin{equation}\label{equ32}
\begin{array}{rcl}
P_{IS2}&=&\sigma_h^2 (|h_{SR}|^2+\sigma_R^2) (\sum\limits_{j=2}^{b+1}|(j-1)\bar{h}_{LI}^{j-2}\beta^{j}|^2+\sum\limits_{j=1}^{b}|\bar{h}_{LI}^{j+b-1}\beta^{j+b+1}|^2)\\
&=&\sigma_h^2 \left(1+\dfrac{\sigma_R^2}{|h_{SR}|^2}\right)\beta^2\left(\sum\limits_{j=1}^{b}|j\bar{h}_{LI}^{j-1}\beta^{j}h_{SR}|^2)+\sum\limits_{j=b+1}^{2b}|\bar{h}_{LI}^{j-1}\beta^{j}h_{SR}|^2\right)\\
&=&\sigma_h^2 \left(1+\dfrac{\sigma_R^2}{|h_{SR}|^2}\right)\beta^4|h_{SR}|^2\frac{1+a^2-2ba^{2b}+3(b-1)a^{2b+2}+(1-b)a^{2b+4}-a^{4b}+2a^{4b+2}-a^{4b+4}}{(1-a^2)^3}
\end{array}
\end{equation}
\hrulefill
\vspace*{4pt}
\end{figure*}

Since $|\Delta h|\ll1$, the contribution of the terms of $\Delta h^p(p\geq2)$ is also neglected as before.
The terms with $\Delta h$ in (\ref{equ30}) are~(\ref{equ31}).
Considering $x(i), i=0,1,2,\ldots$, and $n_R(i), i=0,1,2,\ldots$, are independent, we get the power of the interference  as~(\ref{equ32}),
where $a=|\bar{h}_{LI}\beta|$.

We can control $\beta$ to achieve the minimum upper bound of $P_{IS2}$.
From (\ref{equ3}) and (\ref{equ30}), the SINR at the destination node is obtained as

\begin{equation}\label{equ34-2}
\begin{array}{rcl}
\gamma_{S2}&=&\dfrac{|h_{RD}|^2\beta^2 \sum\limits_{j=1}^{b}|\bar{h}_{LI}\beta|^{2(j-1)}+|h_{SD}|^2}{|h_{RD}|^2\left(\frac{\sigma_R^2}{|h_{SR}|^2}\beta^2\sum\limits_{j=1}^{b}|\bar{h}_{LI}\beta|^{2(j-1)}+P_{IS2}\right)+\sigma_D^2}\\
&=&\dfrac{1+\frac{|h_{SD}|^2}{|h_{RD}|^2}}{\frac{\sigma_R^2}{|h_{SR}|^2}+\Phi(\beta)},
\end{array}
\end{equation}
where
\begin{equation}\label{equ34-3}
\Phi(\beta)=\frac{P_{IS2}(1-|\bar{h}_{LI}{\beta}|^2)}{\beta^2(1-|\bar{h}_{LI}{\beta}|^{2b})}+\frac{\sigma_D^2(1-|\bar{h}_{LI}{\beta}|^{2})}{{|h_{RD}|^2\beta}^2(1-|\bar{h}_{LI}{\beta}|^{2b})}.
\end{equation}

From (\ref{equ34-2}), we notice that $\gamma_{S2}$ can be maximized by minimizing $\Phi(\beta)$. Substituting (\ref{equ32}) into (\ref{equ34-3}),
we obtain (\ref{equ34-4}) at the top of the next page,
where $a=|\bar{h}_{LI}\beta|$. Since $a=|\bar{h}_{LI}\beta|<1$ and $b\geq 2$, (\ref{equ34-4}) can be approximated by the following equation through omitting the terms of $a^n$ whose power $n$ is larger than 2.
\newcounter{MYtempeqncnt}
\begin{figure*}[!t]
\normalsize
\setcounter{MYtempeqncnt}{\value{equation}}
\begin{equation}\label{equ34-4}
\begin{array}{rcl}
\Phi(\beta)&=&\sigma_h^2 \left(1+\frac{\sigma_R^2}{|h_{SR}|^2}\right)\beta^2|h_{SR}|^2\frac{1+a^2-2ba^{2b}+3(b-1)a^{2b+2}+(1-b)a^{2b+4}-a^{4b}+2a^{4b+2}-a^{4b+4}}{(1-a^2)^2(1-a^{2b})}+\frac{\sigma_D^2(1-a^{2})}{{|h_{RD}|^2\beta}^2(1-a^{2b})}
\end{array}
\end{equation}

\setcounter{equation}{39}
\begin{equation}\label{equ34-6}
\beta^*=\mathop{\arg\min}_{0<\beta\leq1}\Phi(\beta)=\mathop{\arg\min}_{0<\beta\leq1}\left\{\sigma_h^2 \left(1+\dfrac{\sigma_R^2}{|h_{SR}|^2}\right)\dfrac{2\beta^2|h_{SR}|^2}{(1-2|\bar{h}_{LI}\beta|^2)}+\dfrac{\sigma_D^2(1-|\bar{h}_{LI}{\beta}|^{2})}{{|h_{RD}|^2\beta}^2}\right\}
\end{equation}
\setcounter{equation}{\value{MYtempeqncnt}}
\hrulefill
\vspace*{4pt}
\end{figure*}

\setcounter{equation}{38}
\begin{equation}\label{equ34-5}
\Phi(\beta)\lesssim\sigma_h^2 \left(1+\dfrac{\sigma_R^2}{|h_{SR}|^2}\right)\dfrac{2\beta^2|h_{SR}|^2}{(1-2|\bar{h}_{LI}\beta|^2)}+\frac{\sigma_D^2(1-|\bar{h}_{LI}{\beta}|^{2})}{{|h_{RD}|^2\beta}^2}.
\end{equation}
Based on this upper bound, the optimal $\beta$ should satisfy (\ref{equ34-6}) at the top of the next page.

\setcounter{equation}{40}
Solving (\ref{equ34-6}), we obtain
\begin{equation}\label{equ34-7}
\beta^*=\sqrt{\frac{1}{\sqrt{\frac{|h_{RD}|^2\sigma_h^2 \left(|h_{SR}|^2+\sigma_R^2\right)}{\sigma_D^2}}+2|\bar{h}_{LI}|^2}}.
\end{equation}
The minimum value of $\Phi(\beta)$ is
\begin{equation}\label{equ34-8}
\Phi_{min}=\frac{4\sigma_D|h_{SR}|^2}{|h_{RD}|}\sqrt{\frac{\sigma_h^2}{|h_{SR}|^2} \left(1+\dfrac{\sigma_R^2}{|h_{SR}|^2}\right)}+\frac{|\bar{h}_{LI}|^2\sigma_D^2}{|h_{RD}|^2}.
\end{equation}

When $\sigma_h^2\gg\sigma_D^2$, the first term in the right
hand side of (\ref{equ34-8}) becomes dominant and we have
$$
\frac{4\sigma_D|h_{SR}|^2}{|h_{RD}|}
\leq \sqrt{\frac{\sigma_h^2}{|h_{SR}|^2} \left(1+\dfrac{\sigma_R^2}{|h_{SR}|^2}\right)},
$$
and
\begin{equation}\label{equ35}
\begin{array}{rcl}
\Phi_{min}&\lesssim&\dfrac{\sigma_h^2}{|h_{SR}|^2} \left(1+\dfrac{\sigma_R^2}{|h_{SR}|^2}\right)+\dfrac{\sigma_D^2}{|h_{RD}|^2}\\
&=&P_{IDD}+\dfrac{\sigma_D^2}{|h_{RD}|^2}.
\end{array}
\end{equation}
Comparing (\ref{equ26-0}) and (\ref{equ34-2}), we obtain $\gamma_{S2}\geq\gamma_{DD}\geq\gamma_{S1}$, where $\gamma_{DD}\geq\gamma_{S1}$ has been proved in Subsection \ref{sec5-b}. This concludes the following result.

If the loop channel error is dominant compared to the noise, i.e., $\sigma_h^2 \gg \sigma_D^2$, then,
$\gamma_{S2}\geq\gamma_{DD}\geq\gamma_{S1}$ can be achieved by controlling the amplifying factor $\beta$ in the proposed Scheme 2.

The physical meaning of controlling the
 amplifying factor $\beta^*$ is to control the relay transmission power according to the loop channel information error and other channel information.
That is, the relay adjusts the transmission power to reduce the interference caused by the loop channel information error.
This idea is first proposed in~\cite{5089955}. The difference between the amplifying factor $\beta^*$ of~(\ref{equ34-7}) and that in~\cite{5089955} comes from two aspects. One is the signal from the direct link is treated as useful signal in this paper while it is regarded as noise in~\cite{5089955}. The other is the signal retransmitted by the relay includes consecutive $b$ symbols in this paper while it includes only one symbol in~\cite{5089955}.
From (\ref{equ34-7}), we notice that the cost we pay for the better SINR is
that the channel information of relay to destination, $h_{RD}$,
is needed at the relay. For a time division duplexing (TDD) system, this
may be obtained by the symmetry of downlink and uplink. For a frequency
division duplexing (FDD) system, this may be obtained by the channel information feedback.

\section{Simulations}\label{sec6}
In this section, we present some simulation results to illustrate the performance of our proposed DLC-STC schemes for full-duplex
cooperative communications. We assume that the direct source to destination channel, source to relay channel, relay to destination channel  and relay loop channel
are all quasi-static Rayleigh flat fading. The delay between the relay and the direct link is uniformly distributed in $[0,\tau_{max}]$.
The length of each information symbol frame is 20. The maximum delay $\tau_{max}$ and zero padding length are both 3. The constellation used is QPSK.
We compare the BER performance vs. SNRs for three schemes: DLC-STC Scheme 1,  DLC-STC Scheme 2, and delay diversity
scheme. All the schemes are evaluated
with  MMSE and/or MMSE-DFE receivers~\cite{869048,5683902}. The signal to noise ratios (SNRs) at the receivers of the relay and the destination are denoted
as $S\!N\!R_R$ and $S\!N\!R_D$, respectively. Since the average power gain of each wireless Rayleigh flat fading channel is normalized to be 1, we have $S\!N\!R_R=\frac{E_s}{\sigma^2_R}$ and $S\!N\!R_D=\frac{E_s}{\sigma^2_D}$.

\begin{figure}[htbp]
\centering
\includegraphics[scale=0.55]{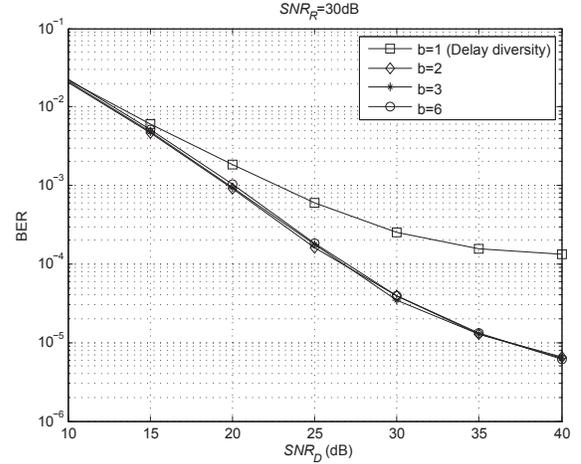}
\caption{\label{fig4-0} Effect of consecutive coding symbols' length $b$ on BER performance of Scheme 2 with MMSE-DFE receivers.}
\end{figure}

\begin{figure}
\centering
\subfigure[BER versus $S\!N\!R_D$]{\label{fig4}\includegraphics[scale=0.55]{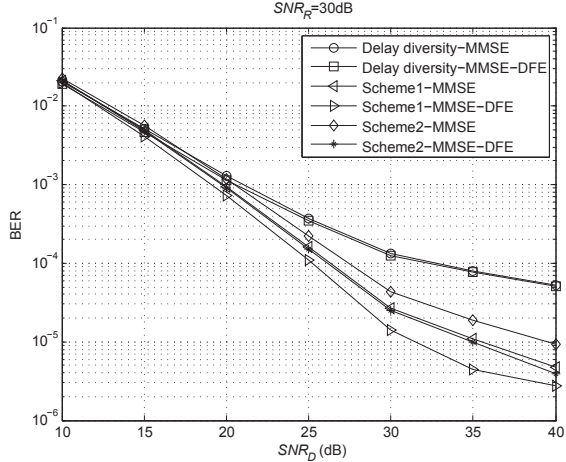}}
\subfigure[BER versus $S\!N\!R_R$]{\label{fig5}\includegraphics[scale=0.55]{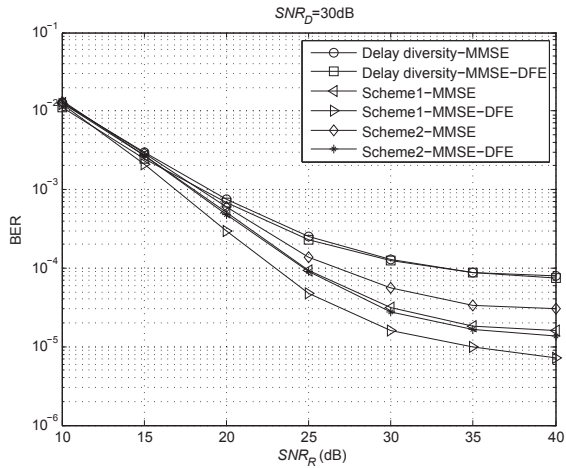}}
\caption{BER performance comparison of different schemes (b=3 for DLC-STC schemes).}
\end{figure}

{\it Simulation 1-Perfect loop channel information}: This simulation is to evaluate the performance of the proposed schemes with loop
channel information known perfectly by the relay. In this simulation, the transmission power at the relay is normalized to be the same in all the schemes. We first investigate the effect of consecutive coding symbols' length $b$ on BER performance of Scheme 2 (To ensure the coding matrix is SFR, $b\geq2$. $b=1$ is the case of delay diversity code.) and the result is shown in Fig.~\ref{fig4-0}
where the receiver is MMSE-DFE. We can see that the performance of Scheme 2 ($b=2$) is better than that of delay diversity code ($b=1$). The difference of the performance is not obvious among the schemes of $b\geq2$. In all the simulations later, we use $b=3$ in Scheme 2. Then we compare the BER performance vs. $S\!N\!R_D$ when $S\!N\!R_R$=30dB and the BER performance vs. $S\!N\!R_R$ when $S\!N\!R_D$=30dB, which are shown in Fig.~\ref{fig4} and Fig.~\ref{fig5}, respectively. From these two figures, we can see that both the proposed DLC-STC schemes outperform the delay diversity scheme and the performance of Scheme 1 is the best. We also notice
that there exist error floors at about 40dB. This is because the noise at the receiver of the relay is also amplified by the relay and transmitted to the destination. When one of the two SNRs is fixed and becomes a bottleneck in the received signal at the destination node, the increase in another will not benefit the BER too much any more.

\begin{figure}[htbp]
\centering
\includegraphics[scale=0.55]{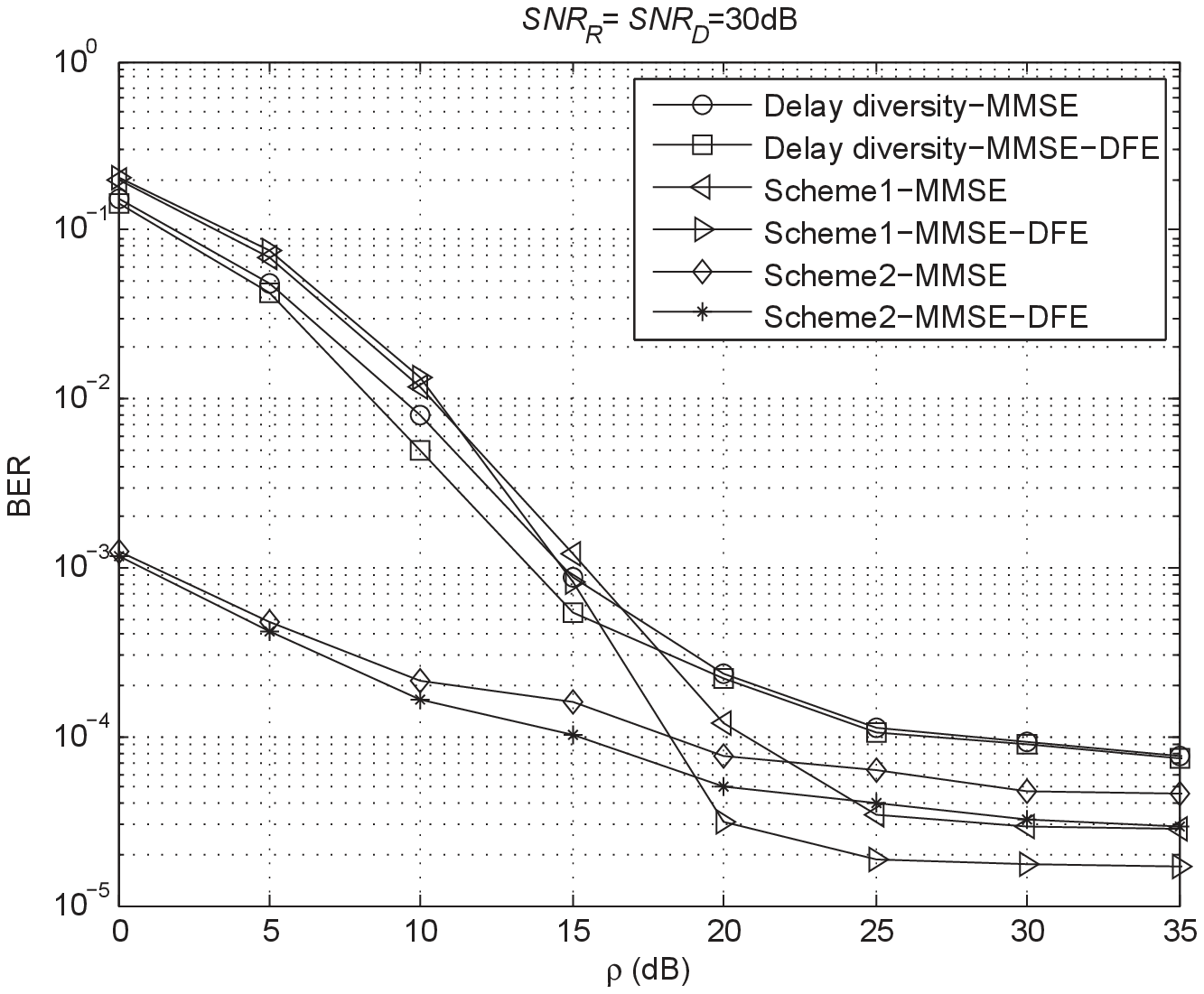}
\caption{\label{fig6} BER versus $\rho$ of full-duplex cooperative communications (b=3 for DLC-STC schemes).}
\end{figure}

{\it Simulation 2-Imperfect loop channel information}: This simulation is to investigate the effects of loop channel information
error on performance of the proposed schemes.
In this simulation, the quality of loop channel information is defined as $\rho=10\log_{10}(\frac{\sigma^2_{{LI}}}{\sigma_h^2})$,
where $\sigma^2_{{LI}}$ is the loop channel power which is assumed to be 1 and $\sigma_h^2$ is the variance of the error. Thus, the
loop channel quality can be written as $\rho=10\log_{10}(\frac{1}{\sigma_h^2})$. In Scheme 2, the amplifying factor $\beta^*$ in~(\ref{equ34-7}) is used. To make sure the error of loop channel information is dominant, $S\!N\!R_R$ and $S\!N\!R_D$ are set to be as large as 30dB in the simulation. Fig.~{\ref{fig6}} shows the BER performance vs. the loop channel information quality $\rho$. We notice that the error of loop channel information degrades the performance of the proposed Scheme 1 the most. The proposed Scheme 2 is much more robust than the other two schemes when the channel information quality is not good. We also notice that when the loop channel information error is becoming comparable with the noise at the destination, the performance of the
proposed Scheme 2 becomes worse. This is because the condition $\sigma_h^2 \gg \sigma_D^2$ is not satisfied, for which the amplifying factor obtained may not be optimal.
If the loop channel information  SNR, $\rho$, is as good as or larger than 19dB, Scheme 1 achieves the best performance.  These results verify
the analysis of the effects of loop channel information errors.

\begin{figure}[htbp]
\centering
\includegraphics[scale=0.55]{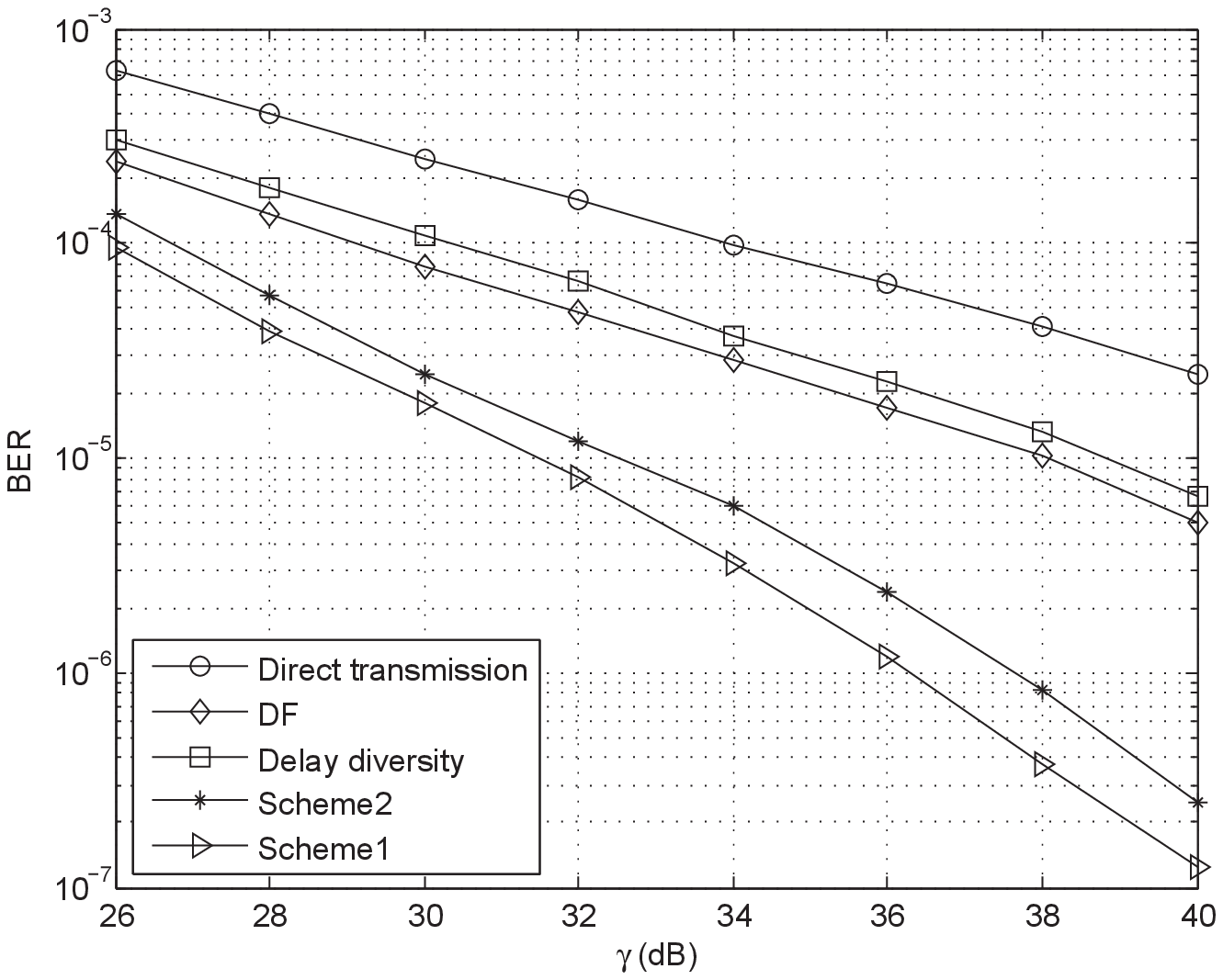}
\caption{\label{fig7} BER versus $\rho$ of full-duplex cooperative communications (b=3 for DLC-STC schemes).}
\end{figure}

{\it Simulation 3-Large SNR}: This simulation is to investigate the achievable diversity when SNR is high. In the simulation, for convenience the loop channel information is assumed to be perfect. And MMSE-DFE receivers are used at the destination node. The SNRs at the relay node and the destination node are set to be equal, that is, $S\!N\!R_R=S\!N\!R_D=\gamma$. Especially, the BER for the reference case with only the direct link called the direct transmission scheme is also included in the simulation. For a fair comparison, the transmission power for direct transmission scheme is doubled, so the SNR is $\frac{2Es}{\sigma_D^2}=2{\gamma}$. This simulation also includes the decoding and forward (DF) scheme
for the delay diversity code for comparison. The results are shown in Fig.~{\ref{fig7}}, where we can see the proposed schemes can achieve the full diversity gain, $2$.

\section{Conclusions}\label{sec7}
In this paper, we presented two DLC-STC schemes for asynchronous full-duplex cooperative communications with direct link. Both schemes can achieve full asynchronous cooperative diversity with MMSE or MMSE-DFE receiver. We showed
 that the proposed Scheme 2 is more robust to the error of loop channel information than the proposed Scheme 1 and delay diversity scheme by controlling the amplifying factor at the relay, when the loop channel information error dominates.

\begin{center}
{\bf Acknowledgement}
\end{center}
The authors would like to thank the anonymous reviewers for their careful reading of this manuscript and for their many detailed, constructive, and
useful comments and suggestions that have improved the presentation of this
paper.

\end{document}